\documentclass[twocolumn,showpacs,amsmath,amssymb]{revtex4}

\usepackage{graphicx}

\newcommand\be{\begin{eqnarray}}
\newcommand\ee{\end{eqnarray}}

\begin{document}
\title{Eliminating nonlinear phase mismatch 
in resonantly enhanced 4-wave mixing}
\author{Mattias Johnsson, Evgeny Korsunsky and Michael Fleischhauer}
\affiliation{Fachbereich Physik, Universit\"{a}t Kaiserslautern, D-67663 Kaiserslautern,
Germany}
\date{\today}

\begin{abstract}
Resonantly enhanced four wave mixing in double-$\Lambda$ systems
is limited by ac-Stark induced nonlinear phase shifts.
With counter-propagating pump
fields the intensity-phase coupling has minimal impact on the dynamics,
but it is of critical importance for co-propagation. 
The nonlinear phase terms lead  to an increase 
of the conversion length linearly proportional to the inverse
seed intensity, while without nonlinear phase-mismatch the scaling
is only logarithmic. 
We here show that the ac-Stark contributions can be
eliminated while retaining the four-wave mixing contribution 
by choosing a suitable five level
system with appropriate detunings. 
\end{abstract}


\maketitle




Ever since the cancellation of resonant
linear absorption and refraction 
via electromagnetically induced transparency (EIT)
\cite{harris1997} was first demonstrated, quantum and nonlinear 
optics have successfully been exploring the
consequences. Many interesting effects have been proposed and
investigated \cite{Marangos}. One important
application of EIT is optical
frequency mixing close to atomic resonances 
where it allows making use of the resonantly enhanced nonlinear
interaction without suffering from linear absorption and refraction.
It has been predicted that EIT could even lead to a new regime of nonlinear 
optics on the level of few light quanta 
\cite{Harris1998,HauHarris99,Imamoglu96}.

\begin{figure}[ht]
\begin{center}
  \includegraphics[width=6cm]{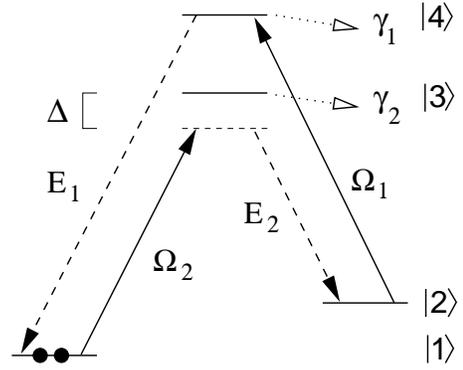}
  \caption{Parametric amplification in a generic double-$\Lambda$ system.}
  \label{fig4level}
 \end{center}
\end{figure}

In this paper we consider one particular EIT-based scheme, namely
resonantly enhanced four wave mixing in a double 
lambda system as shown in Fig.~\ref{fig4level}. 
The two fields with (complex) Rabi frequencies 
$\Omega_1$ and $\Omega_2$ are initially excited and form the pump fields,
while the other fields with (complex) Rabi frequencies $E_1$ and $E_2$ 
are generated during the interaction process. 
$\Omega_1$ and $E_1$ are taken to be exactly on resonance while
the other two are assumed
to be detuned by an amount $\Delta$. 
A finite detuning $\Delta$, large compared to the Rabi frequencies,
Doppler broadening and decay rates from the excited states,
is necessary to maximize the
ratio of nonlinear gain to linear absorption. Decay from the two lower
levels is considered to be negligible. Because of energy conservation
all fields are in four-photon resonance. It can be shown
furthermore
that the contributions of the resonant transitions
to the {\it linear} refractive index vanish if the fields 
are pairwise in two-photon resonance. 
Phase matching will thus favor
two-photon resonance and we assume that this condition is fulfilled.
Resonant four-wave mixing has been analyzed both theoretically and experimentally 
with co-propagating
as well as counter-propagating fields \cite{Hemmer1994,Babin1996,
Lukin1997,Popov1997,Lu1998,Lukin1998,Korsunsky1999,
Fleischhauer00b}.
 
Associated with the finite detuning $\Delta$ are 
ac-Stark shifts which lead to intensity dependent dynamical phase shifts
of the fields. These phase shifts are of minor consequence  
in the case where the fields are counter-propagating
\cite{Fleischhauer00b}. 
They do have a detrimental influence, however,
for co-propagation. 
In the following we will concentrate on the latter situation and
show how to eliminate these terms leading to a 
considerable improvement of nonlinear frequency
conversion.

The standard method to describe the wave mixing process
in a resonant medium is to derive density matrix equations
for the atomic system, solve them in the steady-state, i.e. assuming
adiabaticity,
and insert the resulting expressions into the Maxwell-Bloch equations.
This yields four equations of motion for the fields in the slowly varying
amplitude and phase approximation. The field equations can then be
further broken down into a set of five coupled equations consisting of
four amplitude equations plus the equation of motion governing the
relative phase between the fields.



This procedure can be rather cumbersome, particularly when
several atomic levels are involved. A much simpler way to derive the
field equations is given by the Hamiltonian approach introduced 
in \cite{hamiltonian}, which we will use in the following.
This method makes use of the fact that the polarization $P$ 
of the medium can be expressed as a
partial derivative of the time-averaged interaction energy
per atom $H$ with respect to the electric field
or, equivalently, the Rabi frequencies $E_\mu$
\begin{equation}
P_i=-\frac{N d_i}{\hbar}
\left\langle \frac{\partial {H}}{\partial E_i^*}\right\rangle
\, {\rm e}^{-i(\nu_i (t- z/c)}+c.c.
\end{equation}
A similar expression holds for the drive field polarizations with $E$
replaced by $\Omega$.
Here $\left\langle ...\right\rangle $ denotes quantum-mechanical averaging,
$d_i$ the dipole matrix elements of the corresponding transitions,
and $N$ is the atomic density. Hence introducing moving coordinates
$(\zeta,t)$ with $\zeta=z-ct$
one can directly obtain the field equations of motion
in the slowly-varying amplitude and phase approximation:
\begin{equation}
\frac{{\rm d} E_i}{{\rm d} \zeta}=-i \frac{\eta_\mu}{\hbar}
\left\langle \frac{\partial {H}}{\partial E_i^{\ast }}
\right\rangle ,  \label{Max2}
\end{equation}
where $\eta_\mu = N d_i^2\omega_i/(2\hbar c\epsilon_0)$.
The evaluation of the right hand side of (\ref{Max2}) 
is particularly simple if an open-system model can be used to incorporate
decay in a complex Hamiltonian $H$ and if the atoms adiabatically follow
the dynamics of the fields. If the atoms are initially in an eigenstate
of $H$ with eigenvalue $\lambda$ then 
$\langle H \rangle$ can simply be replaced by $\lambda$ as the two are
equivalent. Thus knowledge of the eigenvalues of the single-atom
interaction Hamiltonian is sufficient to directly derive the field
equations of motion.    

In the basis $(|1\rangle \, |2\rangle \, |3\rangle \, |4\rangle)^T$ the system
shown in Figure~\ref{fig4level} can be described by the complex interaction
Hamiltonian  
\begin{equation}
H=\hbar\left[\begin{matrix}
		0 & 0 & -\Omega_2^* & -E_1^* \cr
                 0 & 0 &  -E_2^*&  -\Omega_1^* \cr
               -\Omega_2 & -E_2 &\Delta - i\gamma_2 &0 \cr
                 -E_1 & -\Omega_1 & 0 & -i\gamma_1 
	\end{matrix}\right].
\label{eqH4level}
\end{equation}
Taking $\Omega$ as a characteristic magnitude of the Rabi frequencies
involved, to second order in $\Omega/\Delta$ the relevant
eigenvalue of (\ref{eqH4level}) is given by
\begin{eqnarray}
\lambda_0 &=& \frac{1}{\Delta} \left[ \frac{\Omega_1 \Omega_2 E_1^*
E_2^* + \Omega_1^* \Omega_2^* E_1 E_2}{|\Omega_1|^2 + |E_1|^2}
\right.\nonumber \\
& &  \qquad \,\, \left. - \frac{|\Omega_1|^2|\Omega_2|^2 + |E_1|^2|E_2|^2}{|\Omega_1|^2 +
|E_1|^2} \right].
 \label{eqlambda0fourlevel}
\end{eqnarray}
This eigenvalue corresponds to the state asymptotically connected to
$|1\rangle$ at $t \rightarrow -\infty$. That is, the eigenstate
associated with $\lambda_0$ corresponds to the ground state $|1\rangle$ for vanishing  
$E_1$ and $E_2$. If the
pump fields change sufficiently slowly we may assume that all atoms
stay at all times in this eigenstate and $\langle H\rangle$ in  (\ref{Max2})
can be replaced by $\lambda_0$. This yields the following equations of motion
\cite{Fleischhauer00b}
\begin{eqnarray}
\frac{\partial}{\partial\zeta} E_1 &=& -i\kappa
\frac{\Omega_1^\ast\Omega_1^2 \Omega_2 E_2^\ast
- E_1^2  E_2  \Omega_1^\ast \Omega_2^\ast}
{\Delta \left(|\Omega_1|^2
+ |E_1|^2\right)^2}
\nonumber\\
&&-i\kappa
\frac{|\Omega_1|^2\left(|\Omega_2|^2-|E_2|^2\right)}
{\Delta \left(|\Omega_1|^2+|E_1|^2\right)^2}\, E_1,\label{eqE1}\\
\frac{\partial}{\partial\zeta} E_2 &=& -i\kappa
\frac{\Omega_1 \Omega_2 E_1^\ast}
{\Delta\left(|\Omega_1|^2+|E_1|^2\right) }
\nonumber\\
&&+i\kappa\frac{
|E_1|^2}
{\Delta \left(|\Omega_1|^2+|E_1|^2\right)}\, E_2,
\end{eqnarray}
\begin{eqnarray}
\frac{\partial}{\partial\zeta} \Omega_1 &=&i\kappa
\frac{\Omega_1^2\Omega_2 E_1^*E_2^*-|E_1|^2E_1 E_2\Omega_2^*}
{\Delta \left(|\Omega_1|^2+|E_1|^2\right)^2}
\nonumber\\
&&+i\kappa\frac{|E_1|^2(|\Omega_2|^2+|E_2|^2)}
{\Delta \left(|\Omega_1|^2+|E_1|^2\right)^2}\,\Omega_1\\
\frac{\partial}{\partial\zeta} \Omega_2 &=& -i\kappa
\frac{E_1 E_2\Omega_1^*}
{\Delta \left(|\Omega_1|^2+|E_1|^2\right)}
\nonumber\\
&&+i\kappa\frac{|\Omega_1|^2}
{\Delta \left(|\Omega_1|^2+|E_1|^2\right)}\, \Omega_2,\label{eqO2}
\end{eqnarray}
where $\kappa=3N\lambda^2\gamma /8 \pi$, with $\gamma$ being 
the decay rate from the upper levels,
where for simplicity we have assumed
$\gamma_1\approx\gamma_2=\gamma$. Note that there are no linear
absorption terms in (\ref{eqE1}) -- (\ref{eqO2}), despite the presence
of the decay terms $\gamma$. Thus, the
process is a parametric one, and the total energy of the
electromagnetic fields is conserved. 
One furthermore finds that the equations have the following three constants
of motion:
\begin{eqnarray}
|\Omega_1|^2 + |E_1|^2 &=& {\rm constant} \\
|\Omega_2|^2 + |E_2|^2 &=& {\rm constant} \\
|\Omega_1|^2 - |\Omega_2|^2 &=& {\rm constant}.
\end{eqnarray}
This allows the problem to be reduced to two degrees of freedom,
one corresponding to the exchange energy between the fields and the
other to the relative phase $\varphi =
\phi_{\Omega 1} + \phi_{\Omega 2} - \phi_{E 1} - \phi_{E
2}$ between the fields.  

The terms in the second line of each equation in (\ref{eqE1}---\ref{eqO2})
are ac-Stark induced, intensity dependent phase terms. 
They have a considerable impact on the dynamics, particularly in
terms of the conversion length, i.e. the distance required for one of the pump
modes to attain maximum transfer into one of the generated modes. To
see this we solve (\ref{eqE1}) --- (\ref{eqO2}) analytically,
using the constants of motion and the methods described in 
\cite{korsunskyET2002,kryzhanovskyET1992}. We will not give
details of the derivation nor the full solution here, however, as they are not 
very instructive for the present purposes.

We consider the case where the two
pump fields are initially of equal intensity, as are the two seed fields, so that
$E_1=E_2=E$ and $\Omega_1 =\Omega_2 =\Omega$. 
If we define the seed field parameter by $\epsilon =
|E(0)|^2/|\Omega(0)|^2$ and denote the initial relative phase
difference between the fields as $\varphi_0$ then, in the limit $\epsilon
\ll 1$, the conversion length $L$ and 
conversion efficiency $e=(|E_{\mathrm{max}}|^2 -
|E_{\mathrm{min}}|^2/(|\Omega_{\mathrm{max}}|^2)$ are given by:
\begin{eqnarray}
e &=& \frac{1-\cos \varphi_0}{1 + \cos \varphi_0}\left[1 -
\epsilon\frac{1-3\cos\varphi_0 -2\cos^2\varphi_0}{1-\cos\varphi_0} \right], \\
L &=& \frac{\Delta}{\kappa} \frac{1}{\sqrt{\epsilon}} \frac{2\pi}{1+\cos\varphi_0}
\end{eqnarray}
One immediately notices that 
 in order to obtain full conversion as $\epsilon \rightarrow 0$
it is necessary to carefully choose the initial phase $\varphi_0$.
Secondly 
the conversion length scales as
$L\sim 1/\sqrt{\epsilon}$, that is, inversely in the seed field amplitude $E(0)$. 
Consequently for small, let alone vacuum, seed
fields, the conversion distance will rapidly become infeasibly
long. 

On the other hand, if the phase terms in  (\ref{eqE1}---\ref{eqO2})
were to be omitted,
the quantity 
\begin{equation}
\mathrm{Re}(\Omega_1 \Omega_2 E_1^* E_2^*) = |\Omega_1 \Omega_2 E_1^*
E_2^*| \cos \varphi,
\end{equation}
is another constant of motion. In this case, if one of the generated fields 
$E_1$, $E_2$ vanishes initially, i.e. if only one of these fields is
seeded, then this constant of motion must be zero. This indicates that the
relative phase $\varphi$ can only jump discontinuously between $\pm
\pi/2$, which occurs only at the end of each conversion cycle, when at least one
of the field amplitudes vanishes. Thus in this case the
phase is essentially decoupled from the evolution of the field amplitudes.
This makes a considerable difference to the dynamics. 
From an analytic solution one finds for the conversion efficiency $e$
and the conversion length $L$
\begin{eqnarray}
e &=& 1-\epsilon\sqrt{\cos\varphi_0}, \\
L &=& \frac{2\Delta}{\kappa}\log \left( \frac{4}{\epsilon^2\cos\varphi_0} \right)
\end{eqnarray}
where we have taken $0<\cos\varphi_0 < 1$ for simplicity.
Thus if the 
phase terms are not present, it is possible to obtain full conversion
in the small seed field limit, regardless of the initial phase
condition.
Secondly, the conversion distance scales only as $-\log
\epsilon$, and the situation is completely 
different to the previous case --- the conversion length will always remain short.

\begin{figure}[ht]
\begin{center}
  \includegraphics[width=7cm]{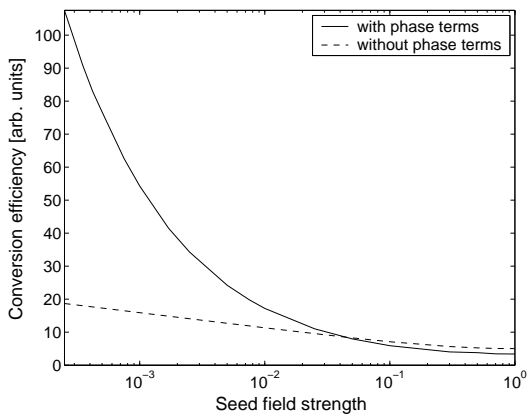}
\caption{Initial field intensities taken as $|\Omega_1(0)|^2 =
|\Omega_2(0)|^2$, $|E_1(0)|^2 = |E_2(0)|^2$, with the seed
field intensity defined as $\epsilon = |E_1|^2/|\Omega_1^2|$.}
\label{figConversionDistances}
\end{center}
\end{figure}

As an illustration of conversion distance dependence on seed field
intensity we have calculated numerical solutions to Eqs. (\ref{eqE1}) --- (\ref{eqO2}) 
with and without the phase terms, and without making the approximation that $\epsilon \ll
1$. We have assumed $E_1=E_2$, $\Omega_1 = \Omega_2$ and $\varphi_0 =
\pi/4$. The results are shown in 
Figure~\ref{figConversionDistances}.

Thus, given the obvious advantages inherent in the omission of these
phase terms, the question naturally arises: does there exist a situation in
which these terms can be made to vanish?

\begin{figure}[ht]
\begin{center}
  \includegraphics[width=6cm]{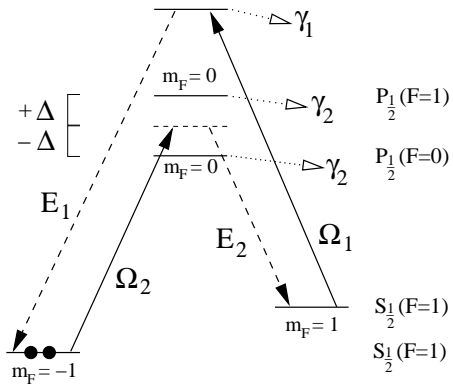}
  \caption{Modified double-$\Lambda$ system. $F$ and $m_F$ represent
the total and $z$-projection angular momentum for the atom, including
hyper-fine interaction.}
\label{fig5level}
\end{center}
\end{figure}

It has been shown by Harris 
\cite{Harris-nl-phase} that in a system of parallel $\lambda$ 
transitions with different excited-state energies there 
exists an optimum detuning such that the 
{\it  nonlinear} index of refraction vanishes.
A similar idea can be applied here.
Noting that both parts of (\ref{eqlambda0fourlevel}) are linear in
$\Delta$, but only the first part is linear in each of the fields,
suggests a method for canceling the phase terms and at the same time
retaining the nonlinear interaction part.
To see this, consider
the five level system shown in Figure \ref{fig5level}.
Here we have introduced hyperfine angular
momentum terms, level $|3\rangle$ in the original scheme
has been split into two and the pump beam $\Omega_2$ tuned to a
point midway between them. Within the approximation that the
electronic parts of the wave functions describing the two $P_{1/2}$ states are
identical, as are the electronic parts of the two $S_{1/2}$ states, we
find that the coupling strengths for the $| S_{1/2} \rangle
\rightarrow |P_{1/2} \rangle$ transitions all have the same magnitude. 
Crucially, however, the
$|S_{1/2}, \, F=1, \, m_F = -1 \rangle \rightarrow | P_{1/2}, \,F=1,
 \, m_F = 0\rangle$ transition has opposite sign to the other three. The
Hamiltonian thus becomes
\begin{equation}
H = -\hbar\left[
   \begin{matrix}
	0 & 0 & \Omega_2^* & \Omega_2^* & E_1^* \cr
        0 & 0 &  E_2^* & -E_2^* & \Omega_1^* \cr
        \Omega_2 & E_2 & -\Delta-i\gamma_2 & 0 & 0 \cr
        \Omega_2 & -E_2 & 0 & \Delta-i\gamma_2 & 0 \cr
 	E_1 & \Omega_1 & 0 & 0 & -i\gamma_1
   \end{matrix}\right].
\label{eqH5level}
\end{equation}
We now find that to second order in $\Omega/ \Delta$ the lowest eigenvalue of (\ref{eqH5level}) is
\begin{equation}
\lambda_0 = \frac{1}{\Delta} \left[ \frac{\Omega_1 \Omega_2 E_1^*
E_2^* + \Omega_1^* \Omega_2^* E_1 E_2}{|\Omega_1|^2 + |E_1|^2}\right].
\label{eqlambda0fivelevel}
\end{equation}
We see that the phase terms
responsible for the increase in the conversion length and sensitive
dependence of the conversion efficiency on the initial phase
are indeed absent. It should be noted that under more general conditions,
e.g. if the dipole moments to the $F=1$ and $F=0$ manifolds are not equal, 
other values of the detunings need to be chosen. 

In summary we have shown that the nonlinear phase contributions arising
in resonant forward four-wave mixing due to the ac-Stark effect can be
exactly eliminated if a five state system with appropriate couplings
and detunings is used. We have derived a simple effective Hamiltonian for
this system and shown that the conversion length scales only
logarithmically with the inverse seed intensity whereas with the phase
terms present conversion length scales linearly.

\def\etal{\textit{et al.}}

\end{document}